\documentclass[conference]{IEEEtran}
\IEEEoverridecommandlockouts
\usepackage{booktabs}

\usepackage{cite}
\usepackage{amsmath,amssymb,amsfonts}
\usepackage{algorithmic}
\usepackage{graphicx}
\usepackage{textcomp}
\usepackage{xcolor}
\usepackage{hyperref}
\usepackage[normalem]{ulem}

\usepackage{multirow}
\usepackage{enumitem, array}
\usepackage{caption}
\usepackage{subcaption}
\usepackage[export]{adjustbox}
\usepackage{hyperref}
\usepackage{gensymb}

\graphicspath{{images/}}

\newcommand{\argmin}{\operatornamewithlimits{argmin }}
\newcommand{\argmax}{\operatornamewithlimits{argmax }}

\def\a{{\bf a}}

\def\f{{\bf f}}

\def\X{{\bf X}}
\def\x{{\bf x}}

\def\v{{\bf v}}
\def\y{{\bf y}}

\def\T{{\bf T}}
\def\t{{\bf t}}

\def\loss{{\mathcal L}}

\def\bphi{{\boldsymbol \phi}}

\def\D{{\mathcal{D}}}

\def\Dtt{{\mathcal{D}_{\mathrm{t} \mathrm{\hat t}}}}
\def\Dt{{\mathcal{D}_{\mathrm{t}}}}

\def\BibTeX{{\rm B\kern-.05em{\sc i\kern-.025em b}\kern-.08em
    T\kern-.1667em\lower.7ex\hbox{E}\kern-.125emX}}

\begin{document}

\title{Machine learning attack on copy detection patterns: are 1x1 patterns cloneable?

\thanks{S. Voloshynovskiy is a corresponding author.}
\thanks{This research was partially funded by the Swiss National Science Foundation SNF No. 200021\_182063.}
}

\author{\IEEEauthorblockN{Roman Chaban, Olga Taran, Joakim Tutt, Taras Holotyak, Slavi Bonev and Slava Voloshynovskiy}
\IEEEauthorblockA{Department of Computer Science, University of Geneva, Switzerland \\
\{roman.chaban, olga.taran, joakim.tutt, taras.holotyak, slavi.bonev, svolos\}@unige.ch}
}
\maketitle

\begin{abstract}

Nowadays, the modern economy critically requires reliable yet cheap protection solutions against product counterfeiting for the mass market. Copy detection patterns (CDP) are considered as such a solution in several applications. It is assumed that being printed at the maximum achievable limit of a printing resolution of an industrial printer with the smallest symbol size $1\times1$, the CDP cannot be copied with sufficient accuracy and thus are unclonable. In this paper, we challenge this hypothesis and consider a copy attack against the CDP based on machine learning. The experimental results based on samples produced on two industrial printers demonstrate that simple detection metrics used in the CDP authentication cannot reliably distinguish the original CDP from their fakes under certain printing conditions. Thus, the paper calls for a need of careful reconsideration of CDP cloneability and search for new authentication techniques and CDP optimization facing the current attack.

\end{abstract}


\IEEEpeerreviewmaketitle

\section{Introduction}

The problem of the security of physical objects and their protection against counterfeiting is among the highly demanded features for the modern society and the economy. The counterfeited products affect numerous life aspects such as health-critical products (medicine, food), luxury products (objects of art, watches), identification documents and banknotes.

Besides the big variety of different anti-counterfeit technologies, all of them have pros and cons and no technique can guarantee perfect protection. For example, holograms are known to be difficult to clone. At the same time, they are not directly suitable for machine authentication and require the users' education and are still quite expensive in production. This makes their usage limited for massive markets. The RFID \cite{10.5555/861917} and chip-based solutions are still expensive, require expensive infrastructure and extra equipment for the verification. One of the popular nowadays technologies is copy detection patterns (CDP) \cite{7472033} \cite{picard2004} that belongs to the group of technologies based on hand-crafted randomness. Moreover, CDP are considered as an efficient yet cheap and user-friendly solution due to their low cost of production. Additionally, the recent achievements of mobile phone technologies allow the products' verification directly by the end customers.

The advantages of CDP also include their easy integration into product design, the wide area of applications and low computational complexity for both enrollment and authentication. Besides the mentioned benefits that make the CDP to be a valuable protection technology, it has been shown in \cite{Taran2019icassp, 10.1145/3335203.3335718} that the CDP might be cloned under certain conditions. However, in the prior works, the authors investigated the cloneability aspects of the CDP printed on the desktop printers in contrast to industrial digital offset printers. In this respect, the current work aims at investigating the cloneability aspect of the CDP printed on the industrial printers under conditions close to those used in the industrial large-scale settings.

\begin{figure}[t!]
    \centering
    \includegraphics[width=0.45\textwidth]{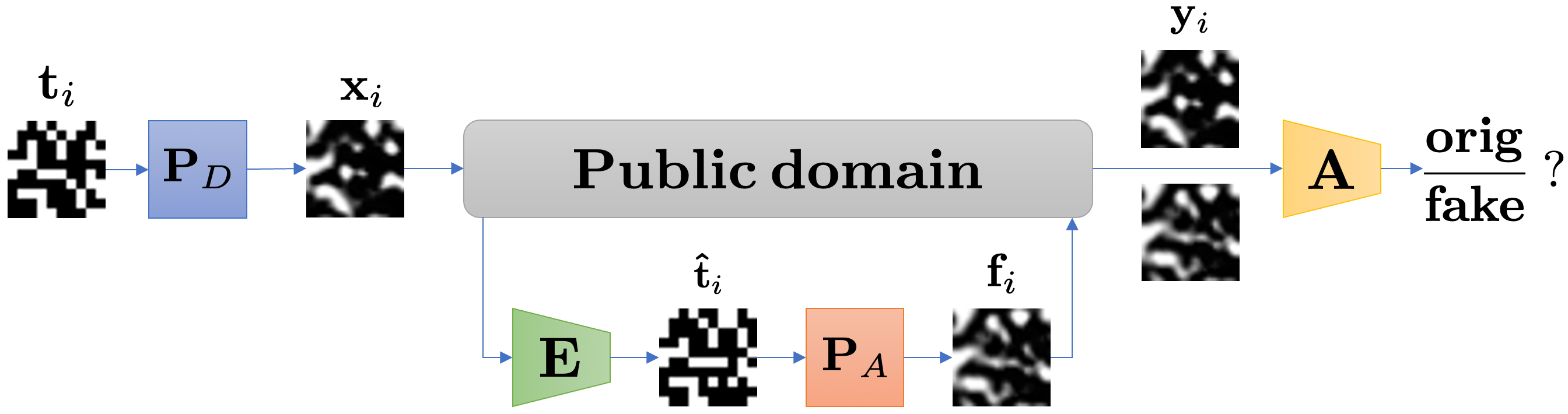}
    \caption{The cycle of the CDP: $\t_i \in \{0, 1\}^{n\times m}$ is a digital template, $P_D$ is the original industrial printer. $E$ is an attacker's estimation network for digital template estimation and $\hat{\t}_i$ is the estimated digital template. $\x_i$ and $\f_i$ are original and fake CDP printed by $P_D$ and $P_A$, respectively. $A$ is the authentication module, which makes a decision whether CDP is original or fake.}
    \label{fig:lifecycle}
\vspace{-4mm}
\end{figure}

The main contributions of the current paper are:

\begin{itemize}
    \item Creation of a public dataset of CDP printed on the industrial printers HP Indigo 5500 and HP Indigo 7600 at the resolution 812.8 dpi with the symbol size $1\times1$.
    \item Investigation of the impact of the different codes' density 30\%, 35\%, 40\%, 45\% and 50\% on the cloneability aspects of the CDP.
    \item Investigation of different authentication methods of the CDP under the produced machine-learning fakes.
\end{itemize}

\begin{figure}[hbt]
    \centering
    \includegraphics[width=0.4\textwidth]{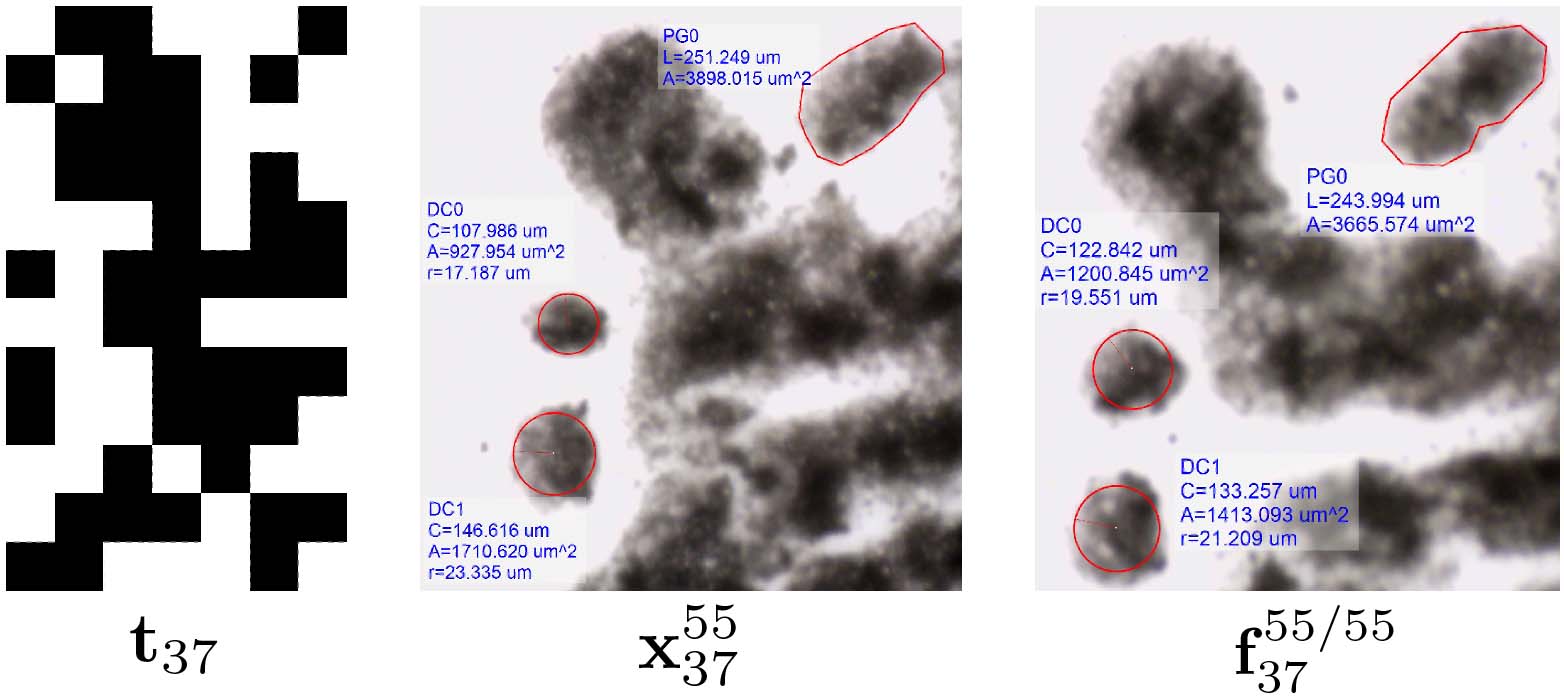}
    \caption[Caption for LOF]{Images of CDP captured with digital microscope\protect\footnotemark. Digital template has symbol size $1\times1$. The subscript denotes the index of CDP, and the superscript denotes the printer used for printing, for the fake $\f$ the second number in superscript denotes the printer used for digital template estimation.}
    \label{fig:dino}
\vspace{-6mm}
\end{figure}

\section{Problem formulation}

As it is shown in Fig. \ref{fig:lifecycle}, the life cycle of CDP starts from the creation of binary digital templates $\t_i \in \{0, 1\}^{n\times m}$, where $n\times m$ denote the size of the code, by the manufacturer and printing them on the industrial printer $P_D$. Printed packages with CDP are distributed in the public domain. An attacker accesses an original package with the CDP, digitizes it and creates a digital fake using a trained estimation model $E$. After that, a fake package with the estimated CDP is printed. In this paper, we consider a possibility that the attacker has an access to a high-quality industrial printing machine. The printed fake package is released to the public domain at any point in time or logistics. At this moment, the authentication algorithm $A$ has to decide whether the received product is original or counterfeited one based on a probe $\y_i$.

In the present work, we assume that the attacker has the same set of knowledge, equipment and algorithms as the defender to implement his/her attack strategies. This includes both access to printing and digitizing equipment. It is a common assumption \cite{picard2004} that a natural obstacle on a way to produce a perfect fake for the attacker is a loss of information in the process of printing due to various factors that are generally referred to as a \textit{dot gain}. To train the CDP estimator $E$, the attacker can create his own training set of digital templates and printed CDP for the model training.

Besides the same access to the equipment, the defender has a freedom in the selection of the authentication algorithm. In this paper, we do not consider the defense by designing a special copy-resistant CDP that looks to be a very attractive defense strategy. Therefore, the defense is only based on a passive authentication. In this work, we investigate both supervised and unsupervised authentication approaches. The supervised authentication might produce more reliable results but needs both originals and fakes for the training of the classifier. On the other hand, a one-class unsupervised authentication requires only a set of originals.

\section{Dataset overview}

\footnotetext{Model: Dino Lite AM7515MT8A (RA), magnification: approximately 689X with made in advance calibration.}

In this paper, we present a new public CDP dataset to investigate CDP cloneability\footnote{The dataset is available \href{https://github.com/sip-group/snf-it-dis/tree/master/datasets/indigo1x1base}{https://github.com/sip-group/snf-it-dis/tree/master/datasets/indigo1x1base}.}. Even though this topic is considered to be quite popular nowadays there is still a lack of datasets produced on industrial printers. This is mainly due to a fact that designing, manufacturing and acquisition of such dataset with further post-processing takes a lot of time and is costly.

\begin{figure}[hbt]
    \centering
    \includegraphics[width=0.45\textwidth]{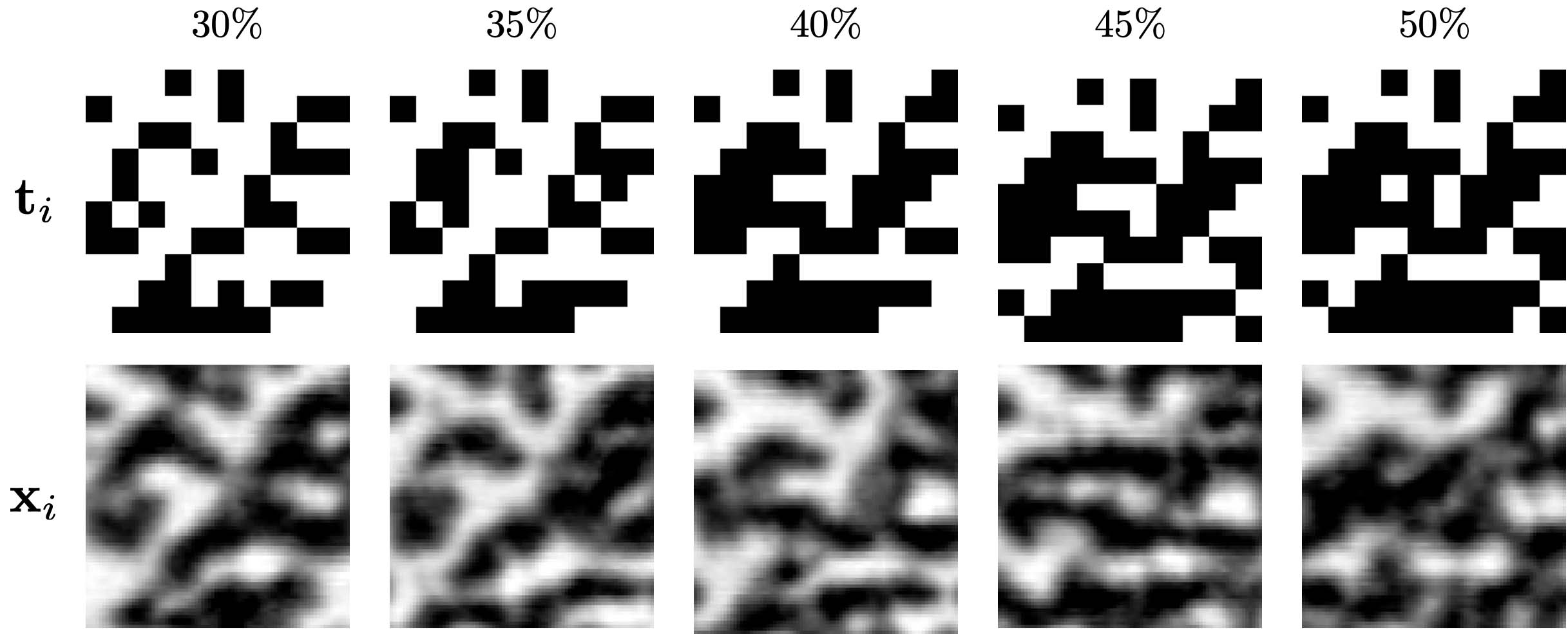}
    \caption{The examples of CDP with different densities. $\t_i$ represents original digital template and $\x_i$ denotes the corresponding printed samples.}
    \label{fig:densities}
\vspace{-4mm}
\end{figure}

\begin{figure}[hbt]
    \centering
    \includegraphics[width=0.45\textwidth]{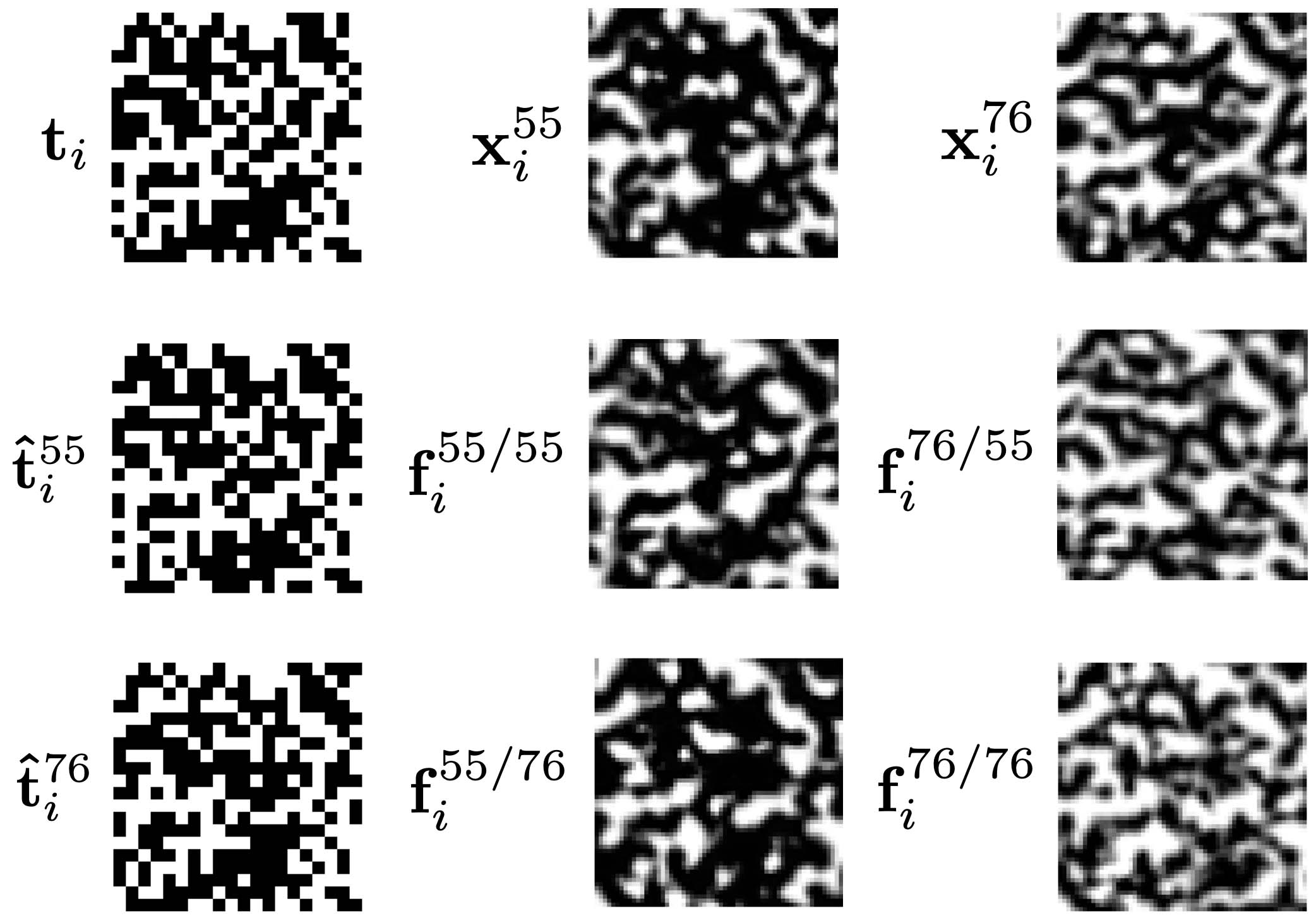}
    \caption{The examples of both digital templates and printed CDP. The digital templates are estimated from printed codes $\x_i^{55}$ and $\x_i^{76}$ and are denoted as $\hat{\t}_i^{55}$ and $\hat{\t}_i^{76}$ respectively. The produced fakes are $\f_i^{55/55}$, $\f_i^{55/76}$, $\f_i^{76/55}$ and $\f_i^{76/76}$, where the first number is a printer on which fake was printed and second number is a printer by which estimation was done.}
    \label{fig:examples}
\vspace{-4mm}
\end{figure}

The produced dataset consists of 720 randomly generated CDP $\t_i \in \{0, 1\}^{228\times228}$. Moreover, to investigate the printer's degradation, we created CDP with different densities. The term density referes to a probability of a black pixel in the CDP, e.g. if $density=30\%$, then the probability of black pixel $P_{bp}=P[\t_{(m,n)}=0]=0.3$. We investigated $30\%, 35\%, 40\%, 45\%, 50\%$ densities. The difference between densities of the same index CDP is shown in Fig. \ref{fig:densities}.

The CDP were printed on two industrial printers HP Indigo 5500 (HPI55) and HP Indigo 7600 (HPI76) with \textbf{812.8} dpi resolution on Invercote G paper. The expected symbol size should be around $30\mu m$ in diameter. The chosen resolution is considered to be the best trade-off between print quality and distortions for the chosen digital offset printers. The overall technical conditions of both printers were not inspected and these factors may lead to certain deviations in printing.

\begin{figure}[hbt]
    \centering
    \includegraphics[width=0.45\textwidth]{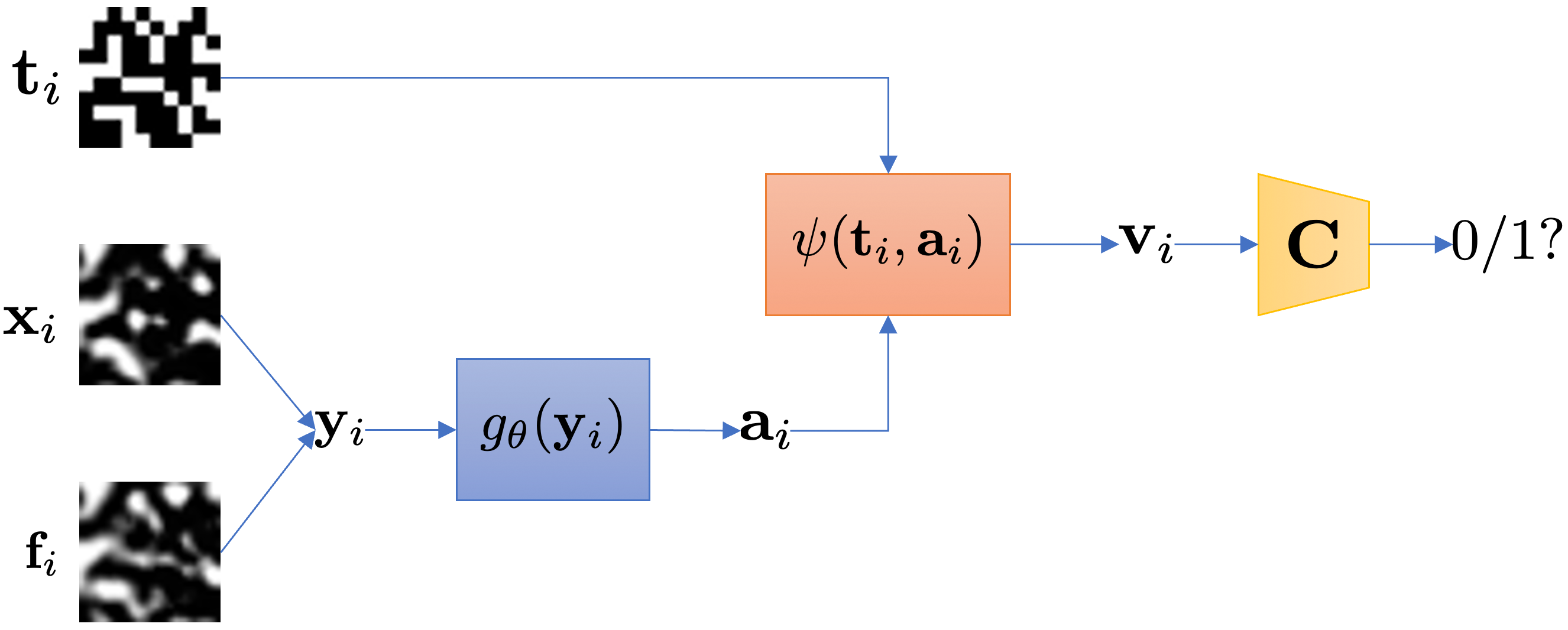}
    \caption{The authentication scheme: $\t_i$ is the digital template. Probe $\y_i$ represents a probe corresponding to $\x_i$ or $\f_i$. $g_\theta(\y_i)$ is the CDP preprocessing function, which consists of synchronization, dynamic range normalization and Otsu's binarization for the Hamming and Jaccard distances and yields a processed image $\a_i$. $\psi(\t_i, \a_i)$ is function or set of functions measuring the differences between the input CDP and digital template. $C$ is the classifier, which can be both one-class and two-class SVM.}
    \label{fig:authentication}
\vspace{-4mm}
\end{figure}

The printed CDP are scanned with Epson Perfection V850 Pro using a particular set of settings\footnote{OS: MacOS 11.3, used software: Epson Scan 2 v6.4.94, unsharp mask: High, brightness = 35, bit depth = 16, color mode = gray.}. The CDP acquisition is performed for the template estimation and CDP authentication. For the CDP estimation considered as a part of fake production, it is important to preserve as much information as possible in the scanned codes and 6400 ppi is the upper optical limit of this scanner model. The acquisition for the authentication stage simulates a hypothetical mobile phone camera resolution which is limited by 2400 ppi. Thus, we used 6400 ppi for the fake production on the side of the attacker and 2400 ppi was used for the authentication to simulate hypothetical mobile phone camera resolution. The resulting digital image of CDP would have about $8\times8$ pixels per each printed symbol with the scanning resolution 6400 ppi and $3\times3$ for 2400 ppi.

Each CDP is placed inside a special design pattern with synchromarkers, which are used for aligning scanned images to digital templates with pixel-wise precision. The visualization of the resulting aligned CDP is shown in Fig. \ref{fig:examples}.

\section{Attack and authentication strategies}

\subsection{Attacking strategy}

The generation of machine learning (ML) fakes is based on an idea of digital template estimation from the printed counterparts. At the second stage, the estimated digital templates are printed on fake packages. In the considered setup, we assumed that besides the publicly available printed codes $\{\x_i\}^M_{i=1}$, the attacker has an access to the corresponding original digital templates $\{\t_i\}^M_{i=1}$. There are various ways to obtain these training pairs. As one possible scenario, one can assume that the attacker can print the digital templates $\{\t_i\}^M_{i=1}$ on the same printer as the defender and scan them as $\{\x_i\}^M_{i=1}$. Such a setup allows to fully explore the power of training on the side of the attacker.

The problem of training an estimator of digital templates from the printed counterparts  given the training data $\{(\t_i, \x_i)\}^M_{i=1}$ generated from a joint distribution $p(\t, \x)$ is formulated as a training of a parameterized network\footnote{The code is available at \href{https://github.com/romaroman/cdp-ml-fakes}{https://github.com/romaroman/cdp-ml-fakes}.} $p_\bphi(\t|\x)$ that is an approximation of $p(\t|\x)$ originating from the chain rule decomposition $p(\t, \x) = p_\D(\x) p(\t|\x)$, where $p_\D(\x)$ corresponds to the empirical data distributions of the original printed codes. The training of the estimation network $p_\bphi(\t|\x)$ is performed based on the maximisation of the mutual information $I_{\bphi}(\T;\X)$ between $\x$ and $\t$ via $p_\bphi(\t|\x)$:

\begin{equation} 
    \hat{\bphi} = \argmax_\bphi I_\bphi(\X;\T) = \argmin_{\bphi}  \loss(\bphi ),
    \label{eq:ml attack max}     
\end{equation}

where $\loss(\bphi ) =  - I_{\bphi}(\T;\X)$.

It was shown in \cite{Voloshynovskiy2019NeurIPS} that the mutual information can be lower bounded as
%
%
$I_\bphi(\X; \T) \ge I_\bphi^L(\X; \T)$, where:
\begin{equation} 
\begin{aligned} 
    I_\bphi^L(\X; \T) 
    & \triangleq \underbrace{\mathbb{E}_{p_\D(\x)} \left[ \mathbb{E}_{p_\bphi(\t|\x)} \left[ \log p_\bphi(\t|\x) \right] \right]}_\text{$\Dtt$} \\
    & \;\;\; - \underbrace{D_{\mathrm{KL}}\left( p_t(\t) \| p_\bphi(\t) \right)}_\text{$\Dt$} \triangleq - \loss^{L}(\bphi), 
\end{aligned}     
\label{eq:dtt dt}  
\end{equation}
where $D_{\mathrm{KL}}\left( p_t(\t) \| p_\bphi(\t) \right) = \mathbb{E}_{p_t(t)} \left[   \log \frac{p_t(\t)}{p_\bphi(\t)}  \right]$ is a Kullback–Leibler divergence between the true $p_t(\t)$ and the posterior $p_\bphi(\t)$.

The final minimization problem reduces to:
\begin{equation} 
\begin{aligned}
    \hat{\bphi} & = \argmin_\bphi \loss^{L}(\bphi)  = \argmin_\bphi  - (\Dtt - \Dt).
\end{aligned}
\label{eq:final optimization problem}
\end{equation}
{\bf Remark:} The term $\Dt$ can be implemented based on the density ratio estimation \cite{GoodfellowGAN}. The term $\Dtt$ can be defined explicitly using Gaussian priors as: $p_\bphi(\t|\x)  \propto \exp(-\lambda\| \t - g_{\bphi}(\x)\|_2)$ with a scale parameter $\lambda$, which leads to $\ell_2$-norm and $g_{\bphi}(\x)$ denotes the parameterized mapper.

The model for digital template estimation $E$ as shown in Fig. \ref{fig:lifecycle} is built on a well-known U-Net architecture \cite{Ronneberger2015} and we refer to it as a template estimation model. The estimator model can be both stochastic and deterministic. The stochastic estimator assumes that the independent additive Gaussian noise with the $std=0.001$ is added to the input before the estimation while the deterministic one does not inject any noise. The results of the obtained estimated and printed CDP are shown in Fig. \ref{fig:examples}.

\subsection{Authentication strategy}

The general scheme of the authentication is shown in Fig. \ref{fig:authentication}. The authentication has to establish whether the probe $\y_i$ representing the original CDP $\x_i$ or fake $\f_i$ is authentic with respect to the template $\t_i$ or not.

\begin{table*}
\centering
\renewcommand*{\arraystretch}{1.2}
\caption{The average probability of error $P_{error}$ in \% of estimation of digital templates from the scanned codes relatively to original template based on normalized Hamming distance. $A_M$ is the attacking method, $P_D$ is the original printer.}
\label{tab:estimation}
\begin{tabular}{ccccccc}
\toprule
     & Density &               30\% &               35\% &                40\% &                45\% &                50\% \\
$A_M$ & $P_D$ &                    &                    &                     &                     &                     \\
\midrule
\multirow{2}{*}{Otsu} & HPI55 &  $12.54\:(\pm0.0)$ &  $14.95\:(\pm0.0)$ &   $16.89\:(\pm0.0)$ &   $18.18\:(\pm0.0)$ &   $20.01\:(\pm0.0)$ \\
     & HPI76 &  $12.88\:(\pm0.0)$ &  $14.97\:(\pm0.0)$ &    $16.5\:(\pm0.0)$ &   $17.61\:(\pm0.0)$ &   $18.13\:(\pm0.0)$ \\
 \cline{1-7}
\multirow{2}{*}{LDA} & HPI55 &  $6.97\:(\pm0.02)$ &  $9.34\:(\pm0.03)$ &   $11.9\:(\pm0.01)$ &  $13.84\:(\pm0.01)$ &  $16.34\:(\pm0.01)$ \\
     & HPI76 &  $7.39\:(\pm0.03)$ &  $9.57\:(\pm0.03)$ &  $11.76\:(\pm0.03)$ &  $13.73\:(\pm0.02)$ &  $15.24\:(\pm0.03)$ \\
\cline{1-7}
$-\Dtt + \Dt^{deter}$ & HPI76 &   $\mathbf{1.68\:(\pm0.1)}$ &   $\mathbf{2.58\:(\pm0.1)}$ &   $\mathbf{3.67\:(\pm0.04)}$ &   $\mathbf{5.12\:(\pm0.15)}$ &   $\mathbf{6.17\:(\pm0.03)}$ \\
$-\Dtt + \Dt^{stoch}$ & HPI55 &  $1.68\:(\pm0.02)$ &  $2.65\:(\pm0.07)$ &   $4.04\:(\pm0.04)$ &   $5.25\:(\pm0.08)$ &   $7.57\:(\pm0.03)$ \\

\bottomrule
\end{tabular}
\vspace{-4mm}
\end{table*}

\begin{figure*}[hbt]
    \centering
    \includegraphics[width=0.92\textwidth]{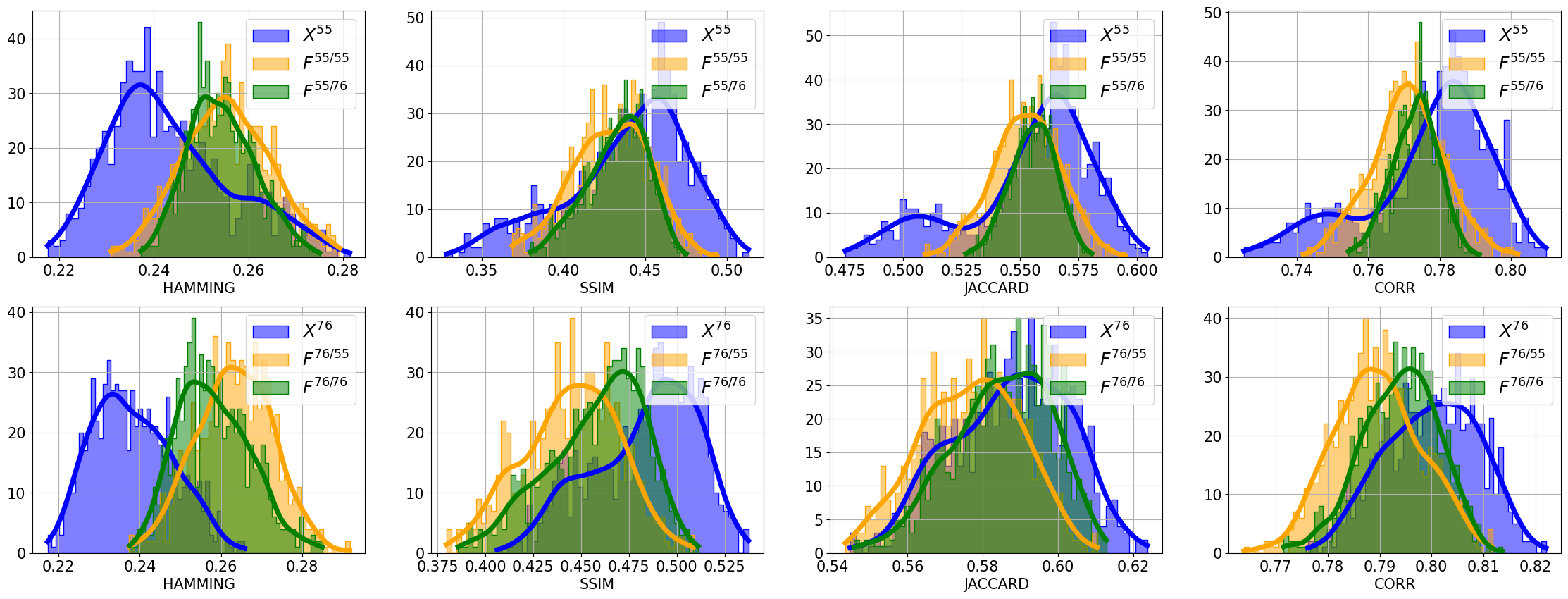}
    \caption{The distributions of metrics and respective approximated kernel density estimation (solid line).}
    \label{fig:dist}
\vspace{-4mm}
\end{figure*}

The probe $\y_i$ is pre-processed to estimate the template $\t_i$ via the mapping $\a_i = g_\theta(\y_i)$. The parameters of mapping $\theta$ can be learnable at the training stage or estimated from data at the moment of the authentication, e.g. like Otsu's thresholding. A similarity metric $\psi(\t_i, \a_i)$ is computed between the template $\t_i$ and the estimate $\a_i$.

\begin{figure}[hbt]
    \centering
    \includegraphics[width=0.48\textwidth]{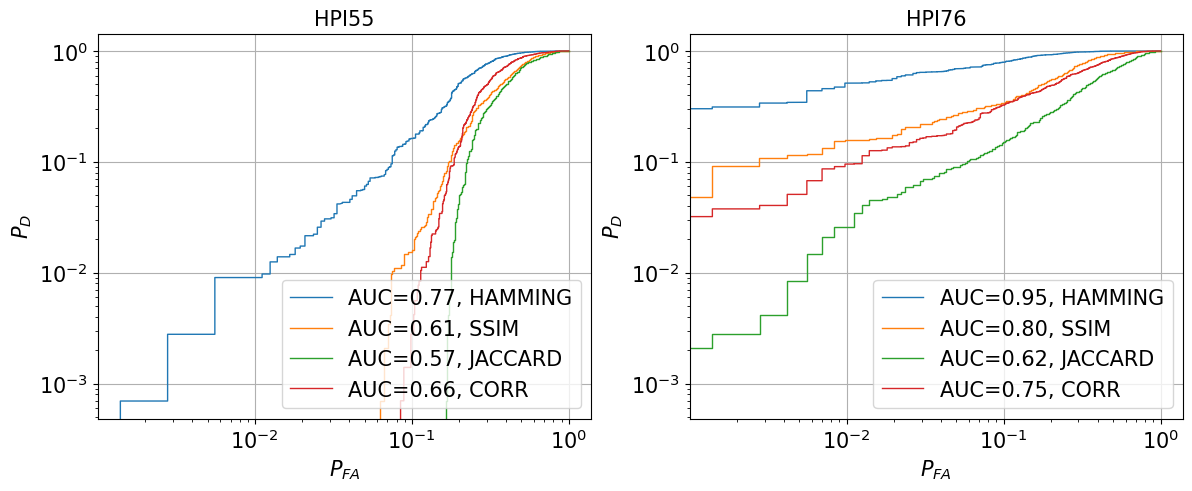}
    \caption{ROC curves with the corresponding area under the curve (AUC) scores for all metrics. The curves for JACCARD, SSIM and CORR are calculated for ${1 - \bf s}$, where ${\bf s}$ is an original score.}
    \label{fig:roc}
\vspace{-6mm}
\end{figure}

The similarity score might include several metrics. In our study, we consider normalized Hamming distance for binary images (HAMMING), structural similarity index (SSIM) \cite{1284395}, Jaccard index (JACCARD) and normalized cross-correlation (CORR). The output vector $\v_i \in \mathbb{R}^4$ is a concatenation of the four above similarity scores. The support vector machine (SVM) classifier $C$ is trained on the vector $\v_i$ to produce a decision about the probe $\y_i$.

\section{Results and discussion}

\subsection{Attack results}

Besides the described template estimation method, we consider two base-line template estimation alternatives based on the LDA algorithm \cite{10.1007/978-3-642-32805-3_4} and binarization based on Otsu's adaptive thresholding \cite{4310076} for benchmarking purposes. The accuracy of template estimation is measured as a normalized Hamming distance between the estimated template $\hat{\t}_i$ and its original counterpart $\t_i$. The template estimation results are presented in Table \ref{tab:estimation} as a corresponding $P_{error}$ which denotes the Hamming distance between binarized estimated CDP $\hat{\t}_i$ and original digital template $\t_i$.

\begin{figure*}[hbt]
    \centering
    \includegraphics[width=0.92\textwidth]{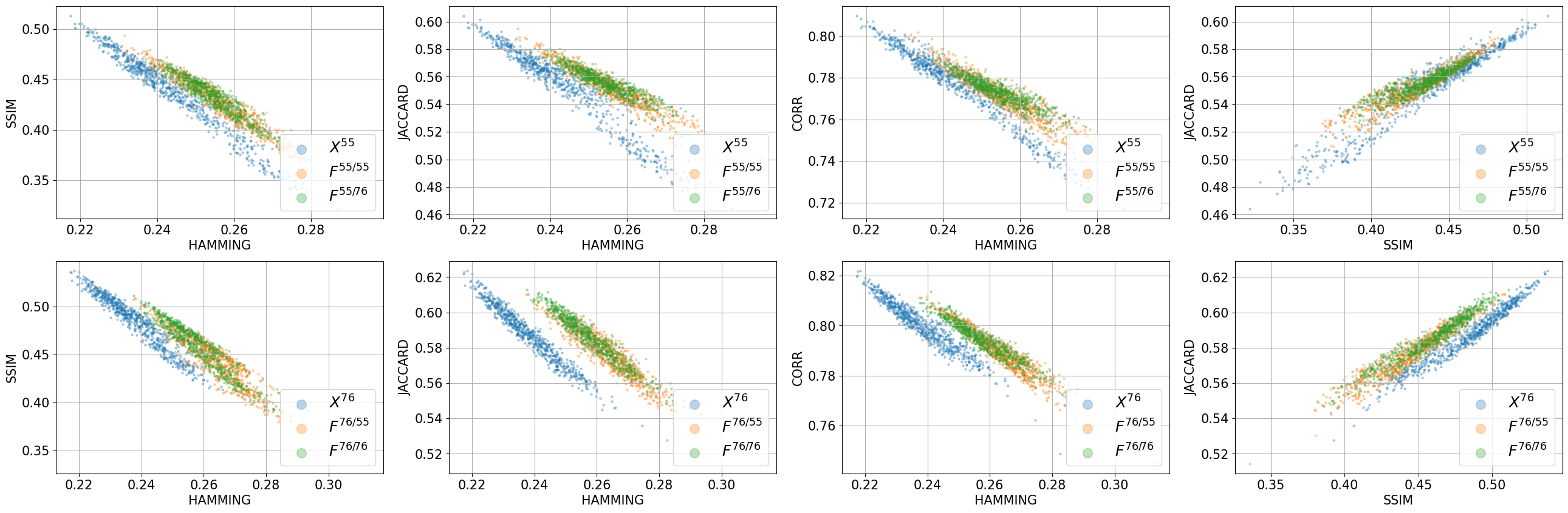}
    \caption{The visualization of pairwise metrics, where each dot represents a particular sample.}
    \label{fig:scatter}
\vspace{-4mm}
\end{figure*}

The comparison between the three estimation techniques indicates that the proposed ML estimation attack produces the highest accuracy for all code densities in comparison to both the Otsu and LDA estimates. Furthermore, the deterministic version of the proposed attack slightly outperforms its stochastic counterpart.

It is important to emphasize the role of code density on attack estimation accuracy. The codes with the highest density of 50\%, i.e., the codes with the highest entropy, are characterized by higher estimation error in comparison to the codes with the lower density.

\begin{table}[t!]
\renewcommand*{\arraystretch}{1.2}
\centering

\caption{The results of the one-class SVM training using pairs of metrics $M_1,\:M_2$. $P_D$ indicates the original CDP printer subset and $P_A$ the fake CDP printer.}
\label{tab:svm1pairs}

\begin{tabular}{ccccc}
\toprule
Metrics & $P_D$ & $P_A$ &  $P_{miss}$ &   $P_{fa}$ \\
\midrule
\multirow{4}{*}{CORR, JACCARD} & \multirow{2}{*}{HPI55} & HPI55 &   \textbf{5.08 (±1.78)} &  \textbf{42.70 (±12.71)} \\
              &       & HPI76 &  31.94 (±7.50) &   82.53 (±3.97) \\
\cline{2-5}
              & \multirow{2}{*}{HPI76} & HPI55 &  42.72 (±6.43) &   11.90 (±6.46) \\
              &       & HPI76 &   \textbf{5.29 (±1.39)} &   \textbf{50.85 (±3.85)} \\
\cline{1-5}
\cline{2-5}
\multirow{4}{*}{HAMMING, SSIM} & \multirow{2}{*}{HPI55} & HPI55 &   \textbf{5.13 (±1.74)} &  \textbf{23.53 (±15.46)} \\
              &       & HPI76 &  93.35 (±5.01) &    0.00 (±0.00) \\
\cline{2-5}
              & \multirow{2}{*}{HPI76} & HPI55 &  99.12 (±0.35) &   61.05 (±6.43) \\
              &       & HPI76 &   \textbf{5.05 (±1.52)} &    \textbf{6.88 (±4.41)} \\
\bottomrule
\end{tabular}

\vspace{-5mm}
\end{table}
Therefore, to address the most challenging task in the investigation of the cloneability of CDP, we proceed with the codes of density 50\%, because we consider that if the authentication system cannot distinguish the fakes with a larger amount of errors, then it will be incapable to distinguish ones with a smaller amount of errors with respect to the original digital template.

At the second stage of the attack, the binary estimates of codes of density 50\% were printed on the same printers HPI55 and HPI76 with the symbol size $1\times1$. The original scans of codes are denoted as $\x^{55}$ and $\x^{76}$ for HPI55 and HPI76, respectively. The fakes of the printed $\x^{55}$ and $\x^{76}$ codes are then reprinted on the same printers and are denoted as $\f^{55/55}$, $\f^{55/76}$, $\f^{76/55}$, $\f^{76/76}$. One can observe some minor differences between printed codes produced by two printers in Fig. \ref{fig:dino}. At this stage, we were not able to establish a source of these differences due to the limited access to the tuning of the printers and not knowing the history of their prior usage.

The examples of the original codes and their fakes are shown in Fig. \ref{fig:dino} and Fig. \ref{fig:examples}. Fig. \ref{fig:examples} presents the results scanned by the scanner at the resolution 2400 ppi, while Fig. \ref{fig:dino} shows the magnified prints acquired by Dino microscope. As one can conclude based on the visual inspection of images that the produced fakes very closely resemble the original CDP. Moreover, the dot size of fakes and originals is almost identical. However, it is obvious, that even there are some minor noticeable differences between the originals and fakes under the microscope examination, it is very important to verify, whether such differences can be reliably authenticated under the mobile phone or even scanner acquisition. The next section validates this question.
\begin{table}
\centering
\renewcommand*{\arraystretch}{1.2}

\caption{The results of the one-class SVM training using all metrics. $P_D$ indicates the printer used for the originals and $P_A$ is the attacker's printer. Both $P_{miss}$ and $P_{fa}$ are calculated over the test subset $\x^{55}, \f^{55/55}$, $\f^{55/76}$ for $P_A$ HPI55 and $\x^{76}, \f^{76/55}$, $\f^{76/76}$ for $P_A$ HPI76.}
\label{tab:svm1all}

\begin{tabular}{ccccc}
\toprule
$P_D$ & $P_A$ & $P_{miss}$ &       $P_{fa}$ \\
\midrule
\multirow{2}{*}{HPI55} & HPI55 &   \textbf{10.91 (±2.28)} &  \textbf{26.39 (±8.90)} \\
      & HPI76 &   99.76 (±0.15) &   0.00 (±0.00) \\
\cline{1-4}
\multirow{2}{*}{HPI76} & HPI55 &  100.00 (±0.00) &   1.28 (±0.40) \\
      & HPI76 &    \textbf{9.92 (±1.86)} &   \textbf{0.00 (±0.00)} \\
\bottomrule
\end{tabular}

\vspace{-5mm}
\end{table}

\subsection{Authentication results}

In this section, we investigate the performance of the authentication system represented in Fig. \ref{fig:authentication}. The authentication is analyzed for the original CDP of density 50\% printed on both printers and the fakes considered in the previous section. The performance validation includes three stages. First, we analyze the discrimination in each of the considered metrics $\psi(\t_i, \a_{i})$ by plotting the corresponding histograms and receiver operating characteristic (ROC) curves. Second, we investigate the classification under the absence of information about the fakes based on a one-class classifier. Third, we consider the classification in the fully informed setting based on a supervised classifier.

The discrimination in different metrics is shown in Fig. \ref{fig:dist}. We plot the distribution of resulting scores between the reference templates $\t_i$ and original prints $\x^{55}$ versus the scores between $\t_i$ and fakes $\f^{55/55}$ and $\f^{55/76}$ and the same is done for HPI76. In all considered metrics, the fakes produced on the same printer as the original CDP are closer to the originals than the cross-printed ones\footnote{It should be noted that in the case of the distribution of the original codes (blue histogram) printed on the HPI55 printer one can observe a small hump that deviates from the main peak. It is related to the instability in the printing quality. This is a very important factor that might seriously impact the CDP authentication and should be properly addressed in the future investigations.}.

The ROC curves produced for two printers HPI55 and HPI76 and the most similar fakes $\f^{55/55}$ and $\f^{76/76}$ are shown in Fig. \ref{fig:roc}. Although none of the considered metrics allows to reliably detect fakes, the Hamming distance between $\t_i$ and binarized $\y_i$ based on the Otsu's thresholding (HAMMING) produces the best results for both printers.

Since none of the considered metrics is capable to achieve satisfactory authentication accuracy of considered CDP under the proposed ML attack, we investigate a simultaneous usage of several scores. Fig. \ref{fig:scatter} shows a pair-wise separability of originals and fakes. The most promising separability plots are observed for the pair of \textit{HAMMING} and \textit{CORR}. For the printer HPI76 the set of original $\x^{76}$ is visually separable from sets of $\f^{76/55}$ and $\f^{76/76}$ fakes. However, such a phenomenon is not observed for the HPI55 printer.

Furthermore, before the training of SVM models, it is important to decide whether to continue with a whole set of metrics or with the best separable ones. To confirm this hypothesis the one-class SVM was trained\footnote{Training set size = 144 samples, number of runs with different seeds = 20, $\gamma=0.3$, kernel = RBF, for one-class SVM $\nu=0.01$.} on the subset of 2 selected metrics, training subset consists only of original samples printed on a single printer (HPI55 or HPI76). The results are presented in Table \ref{tab:svm1pairs} with $P_{miss}$ denoting the probability of classifying the original CDP as fake and $P_{fa}$ the probability of classifying the fake CDP as original. The obtained results confirm that the one-class SVM trained on the target printer used to print the original CDP produces the best results for the corresponding samples. However, $P_{miss}$ drastically increases, when the one-class SVM is used to the wrong printer. 

\begin{table}
\centering
\renewcommand*{\arraystretch}{1.2}
\caption{The results of the supervised two-class SVM training using all metrics. $P_D$ indicates the printer of the training subset and $P_A$ the printer of the test subset.}
\label{tab:svm2all}

\begin{tabular}{ccccc}
\toprule
Trained on & $P_D$ & $P_A$ & $P_{miss}$ &        $P_{fa}$  \\
\midrule
\multirow{2}{*}{$x^{55}$, $f^{55/55}$} & \multirow{2}{*}{HPI55} & HPI55 &   \textbf{2.87 (±0.63)} &    \textbf{2.25 (±0.71)} \\
                      &       & HPI76 &  47.73 (±5.60) &    3.86 (±2.41) \\
\cline{1-5}
\cline{2-5}
\multirow{2}{*}{$x^{55}$, $f^{55/76}$} & \multirow{2}{*}{HPI55} & HPI55 &   \textbf{2.16 (±0.58)} &    \textbf{5.24 (±0.80)} \\
                      &       & HPI76 &  27.54 (±2.86) &   22.59 (±5.14) \\
\cline{1-5}
\cline{2-5}
\multirow{2}{*}{$x^{76}$, $f^{76/55}$} & \multirow{2}{*}{HPI76} & HPI55 &  9.99 (±10.68) &  79.81 (±14.86) \\
                      &       & HPI76 &   \textbf{0.42 (±0.36)} &    \textbf{0.35 (±0.23)} \\
\cline{1-5}
\cline{2-5}
\multirow{2}{*}{$x^{76}$, $f^{76/76}$} & \multirow{2}{*}{HPI76} & HPI55 &   5.12 (±9.50) &  90.39 (±14.43) \\
                      &       & HPI76 &   \textbf{0.20 (±0.34)} &    \textbf{0.22 (±0.23)} \\
\bottomrule
\end{tabular}

\vspace{-2mm}
\end{table}
\begin{figure}
    \centering
    \includegraphics[width=0.48\textwidth]{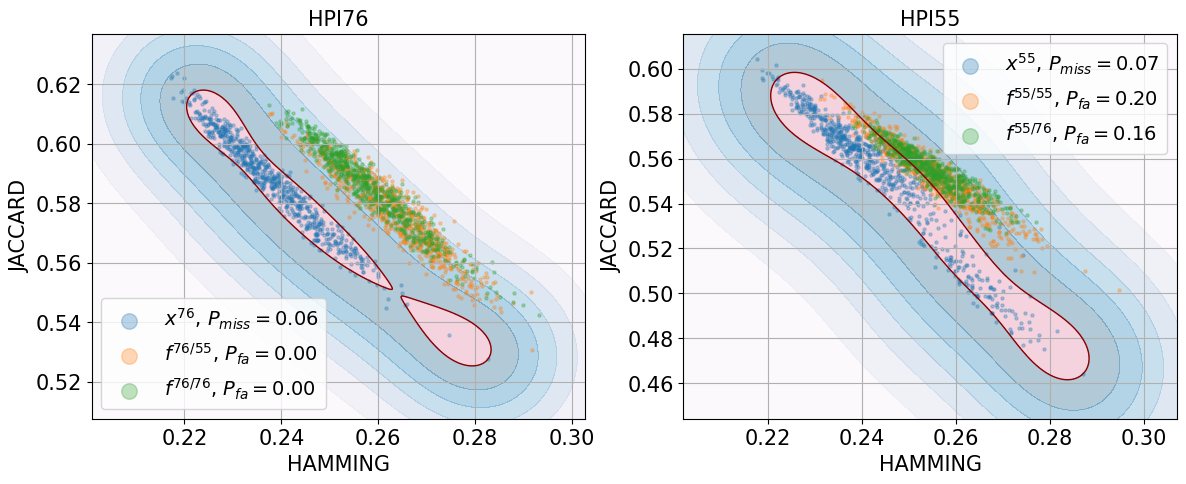}
    \caption{The visualization of one-class SVM decision regions.}
    \label{fig:svm1}
\vspace{-5mm}
\end{figure}

The obtained results indicate that even if the authentication is performed on the digital scanner with 2400 ppi that is generally superior to the mobile phone acquisition, the reliable authentication of CDP is questionable under the proposed ML attack in the one-class SVM setup. The examples of corresponding decision regions are shown in Fig. \ref{fig:svm1}.

To benefit from a full power of SVM the models were trained once again based on the whole set of metrics with the same hyperparameters and the results are present in Table \ref{tab:svm1all}. The full set of metrics does not improve overall $P_{miss}$ but achieves $P_{fa}=0$ for HPI76, while $P_{fa}$ for HPI55 is still high. These results are not satisfactory for practical applications since many originals will be recognized as fakes.

Finally, in the same manner, two-class supervised SVM models were trained as a fully informed system and the results are shown in Table \ref{tab:svm2all}. As expected the supervised approach yields much better $P_{miss}$ and $P_{fa}$. In the best scenario, if SVM is trained on $\x^{76}$ and $\f^{76/76}$, then $P_{miss} =0.20$ and $P_{fa}=0.22$. Still, this result is not satisfactory for the large-scale industrial application and demands to have an access to a big variety of fakes. However,  due to rapid technological growth one cannot guarantee that the classifier will be aware of all unseen fakes and possible attacks. Nevertheless, the trend of inferior accuracy for HPI55 printer is preserved even for the two-class supervised SVM.

\section{Conclusions}
In this work we investigated the possibility to clone CDP with different codes densities produced on two industrial HP Indigo printers.

To perform attacking with the highest possible success chance we assume that the attacker possesses the same equipment and has an access to the original digital templates. This work proves that it is possible to obtain estimations with a relatively low probability of bit error over the whole set of code densities.

The faked CDP were acquired and processed in the same way as originals. The proposed classification system shows that high-quality fakes still preserve some information loss with respect to the originals and can be differentiated. However, the accuracy of authentication should be considerably enhanced for large-scale applications\footnote{Moreover, the performed experiments raise an important question of the impact of the printing variability on authentication accuracy. We aim at investigating this question in more detail in our future research.}.

For future work, we aim at replacing considered SVM classifiers with deep neural networks. Also, it is important to investigate an opportunity to perform authentication without access to the original digital template. Finally, for a complete real-world setup simulation, it is important to acquire present CDP with the mobile phone whereas in this paper it is done on the images acquired by the high-resolution scanner.

\bibliographystyle{IEEEtran}

\bibliography{IEEEexample}

\end{document}